\documentclass[12pt]{iopart} 

\usepackage{subfig}

\expandafter\let\csname equation*\endcsname\relax
\expandafter\let\csname endequation*\endcsname\relax
\usepackage{amsmath}
\usepackage{graphicx}
\usepackage{multirow}

\usepackage{iopams,cite}
\input xy
\xyoption{all}
\usepackage{amsthm}
\usepackage{amsfonts,amsmath}
\usepackage{lscape}

\def\d{\dagger}

\newcommand\phup{^{\phantom p}}

\theoremstyle{definition}

\theoremstyle{remark}

\begin{document}
\title[BCS model with asymmetric pair scattering]{BCS model with asymmetric pair scattering: a non-Hermitian, exactly solvable Hamiltonian exhibiting generalised exclusion statistics}
\author{
Jon Links, Amir Moghaddam, and Yao-Zhong Zhang}
\address{Centre for Mathematical Physics, School of Mathematics and Physics, \\
The University of Queensland 4072,
 Australia}

\eads{\mailto{jrl@maths.uq.edu.au}}

\date{}


\begin{abstract}
\noindent We demonstrate the occurrence of free quasi-particle excitations obeying generalised exclusion statistics in a BCS model with asymmetric pair scattering. The results are derived from an exact solution of the Hamiltonian, which was obtained via the algebraic Bethe ansatz utilising the representation theory of an underlying Yangian algebra. The free quasi-particle excitations are associated to highest-weight states of the Yangian algebra, corresponding to a class of analytic solutions of the Bethe ansatz equations. 
\end{abstract}

\pacs{74.20.Fg, 03.65.Fd, 05.30.Pr}

\section{Introduction}
In a recent Fast Track Communication \cite{lmz12} we discussed the phenonemon of generalised exclusion statistics \cite{h91,i94,p96} in a non-Hermitian BCS model. The non-Hermitian Hamiltonian results from an asymmetry in the Cooper pair scattering amplitudes, depending on whether the scattering is from high energy to low energy states or vice versa. As a result of being non-Hermitian, the Hamiltonian admits complex eigenvalues for particular choices of the coupling parameters. However as the Hamiltonian is a real-valued operator non-real eigenvalues, when they occur, arise as complex conjugate pairs. In this sense the spectrum of the Hamiltonian mirrors the property of admitting an unbroken ${\mathcal PT}$-symmetric phase and a broken ${\mathcal PT}$-symmetric phase which has been discussed in multiple other models, e.g. see articles contained within the Journal of Physics A Special Issue {\it Quantum physics with non-Hertian operators} \cite{bfgj12}. It was found from numerical diagonalisation of the asymmetric BCS Hamiltonian that the transition between broken and unbroken ${\mathcal PT}$-symmetry occurs on two boundary lines in the phase diagram. Furthermore, it was determined from the exact Bethe ansatz solution of the Hamiltonian that there exist free quasiparticle excitations on the boundary lines which can be characterised as satisfying generalised exclusion statistics. This finding relies on a deep understanding of the representation theory of a Yangian algebra, denoted $Y[gl(2)]$, from which the model may be obtained via the Quantum Inverse Scattering Method \cite{dl04}. The objective of this article is to provide a detailed exposition of these generalised exclusion statistics.    

In Section 2 we begin by discussing the reduced BCS pairing Hamiltonian in general terms. We specify the asymmetric choice of Cooper pair scattering amplitudes which leads to our model of investigation. In Section 3 we review some relevant aspects of the Quantum Inverse Scattering Method \cite{ks79,tf79}, and the associated techniques of the algebraic Bethe ansatz for the derivation of exact solutions. This algebraic approach is essentially an application of the representation theory for the Yangian algebra $Y[gl(2)]$, which was developed by Chari and Pressley \cite{cp91}. In Section 4 we illustrate how excitations  obeying generalised exclusion statistics are identified in terms of dividing the solutions of the Bethe ansatz equations into two classes. This situation is reminiscent of generalised exclusuion statistics in an exactly solved model studied by Fendley and Schoutens \cite{fs07}, and provides an example of deconfined quantum criticality \cite{sbsvf04,svbsf04}. In part our analysis amounts to addressing an old question regarding the completeness of the Bethe ansatz solution discussed in Bethe's original work for the $XXX$ chain \cite{b31}, and more recently debated in \cite{b02,fm01}.  We conjecture a formula for counting the number of deconfined states, and prove its validity in some limiting cases. Concluding remarks are given in Section 5. Two Appendices are included. The first tables some complete, analytic solutions of the Bethe ansatz equations associated with low-dimensional Hilbert spaces of states. The second describes a mean-field analysis of the Hamiltonian which predicts that the Hamiltonian always has a real eigenspectrum. These latter calculations highlight the need for exact results to accurately characterise the system.


\section{The Hamiltonian}

The general form for a reduced BCS Hamiltonian as originally discussed in \cite{bcs57} is given by 
\begin{align} H_{\rm{BCS}}=\sum_{j=1}^{L}\epsilon_j n_j
-\sum_{j,k=1}^{L}G_{jk}\phup 
c_{k+}^{\d}c_{k-}^{\d}c_{j-}\phup c_{j+}\phup. \label{bcs_eqn} 
\end{align}
Here, $j{=}1,\dots,{L}$ labels a shell of doubly degenerate single
particle
energy levels with energies $\epsilon_j$, and $n_j=c^\dagger_{j+}c_{j+}\phup
 + \ c^\dagger_{j-}c_{j-}\phup $ is the
fermion number operator for
level $j$. The operators $c_{j\pm}\phup ,\,c^{\d}_{j\pm}$ are the annihilation
and creation operators for fermions at level $j$. The labels $\pm$
refer
to pairs of time-reversed states.

An important feature of the Hamiltonian (\ref{bcs_eqn})
is the blocking
effect. For any unpaired fermion at level $j$ the action of
the pairing interaction is zero since only paired fermions are
scattered. This means that basis states for the Hilbert space can be decoupled into
products of paired and unpaired fermion states in which the
action of the Hamiltonian on the space for the unpaired fermions is
automatically diagonal in the basis of number operator eigenstates.
In view of this property the pair number operator 
$$N=\sum_{j=1}^L c^\dagger_{j+}c_{j+}c_{j-}^\dagger c_{j-}$$
commutes with (\ref{bcs_eqn}) and thus provides a good quantum number. Below, $M$ will be used to denote the eigenvalues of the pair number operator.


It is convenient to express the Hamiltonian in terms of realisations 
of $L$ copies of the $su(2)$ algebra in the pseudo-spin representation,
through the identification
\begin{align} S_j^-=c_{j-}\phup c_{j+} , \quad  \quad 
S_j^+=c^{\d}_{j+}c^{\d}_{j-}, \quad  \quad
S_j^z=\frac{1}{2}\left(n_j -I\right). \label{psr} 
\end{align}
The pseudo-spin operators satisfy the following $su(2)$ commutation relations: 
\begin{align}
[S_j^z, S_k^{\pm}] = \pm \delta_{jk} S_j^\pm, ~~~~[S_j^+,S_k^-] =2\delta_{jk} S_j^z.\label{su2} 
\end{align}
Through this correspondence it is possible to identify a particular form of the Hamiltonian (\ref{bcs_eqn}) as an exactly-solvable model associated with the $su(2)$-invariant six-vertex solution of the Yang-Baxter equation \cite{dl04}. This occurs for the choice  
\begin{align}
G_{jk}=\begin{cases}
  G_+ & j< k \\
  \displaystyle \frac{G_++G_-}{2} & j=k \\
  G_- & j>k
\end{cases} 
\label{gees}
\end{align}
for arbitrary $G_+$ and $G_-$. For real-valued $G_+=G_-$ this model corresponds to the case examined long ago by Richardson \cite{r63}. In \cite{dl04} the specific case where $G_+$ and $G_-$ are a complex conjugate pair, leading to a self-adjoint Hamiltonian, was recognised as the {\it Russian Doll} BCS model of \cite{lrs04} (see also \cite{als05}). 

\begin{figure}[ht!]
    \label{fig:subfigures}
    \begin{center}
        \subfloat[]{  \label{fig:first}  \includegraphics[width=0.4\textwidth]{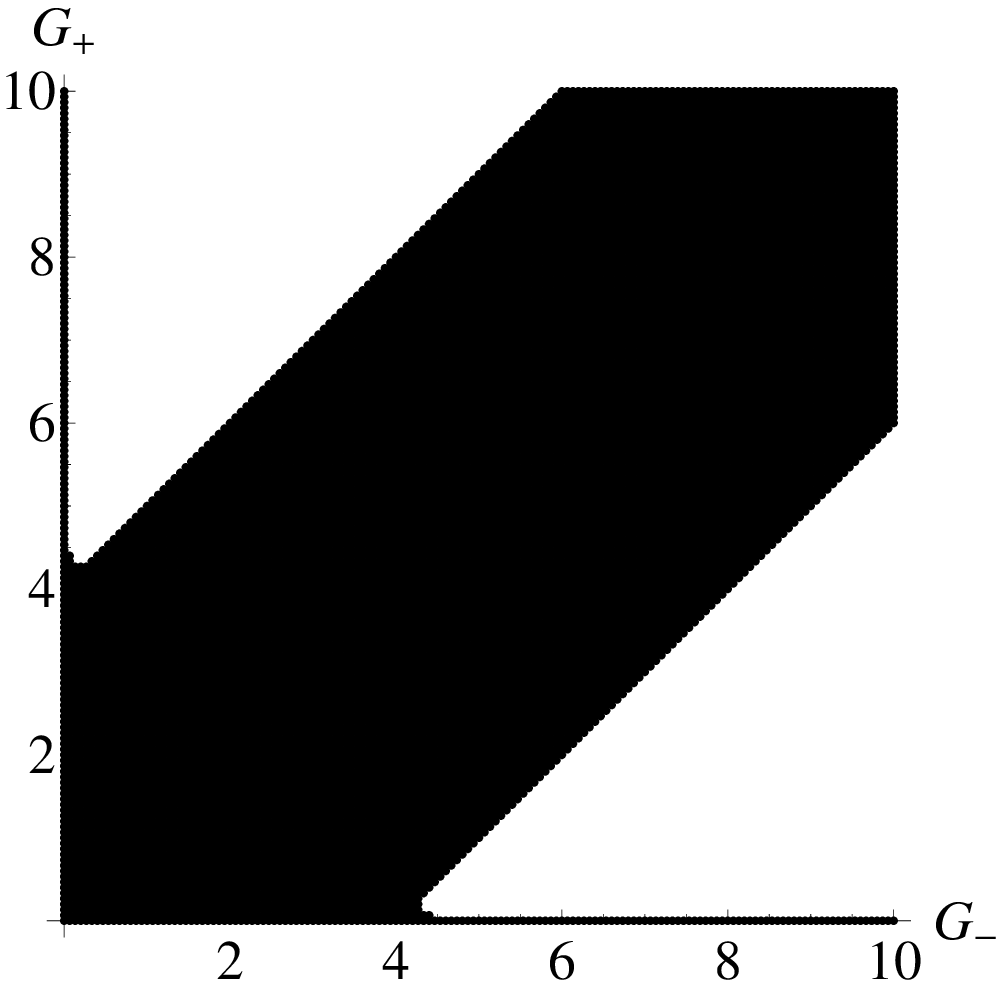}  }
        \subfloat[]{  \label{fig:second} \includegraphics[width=0.4\textwidth]{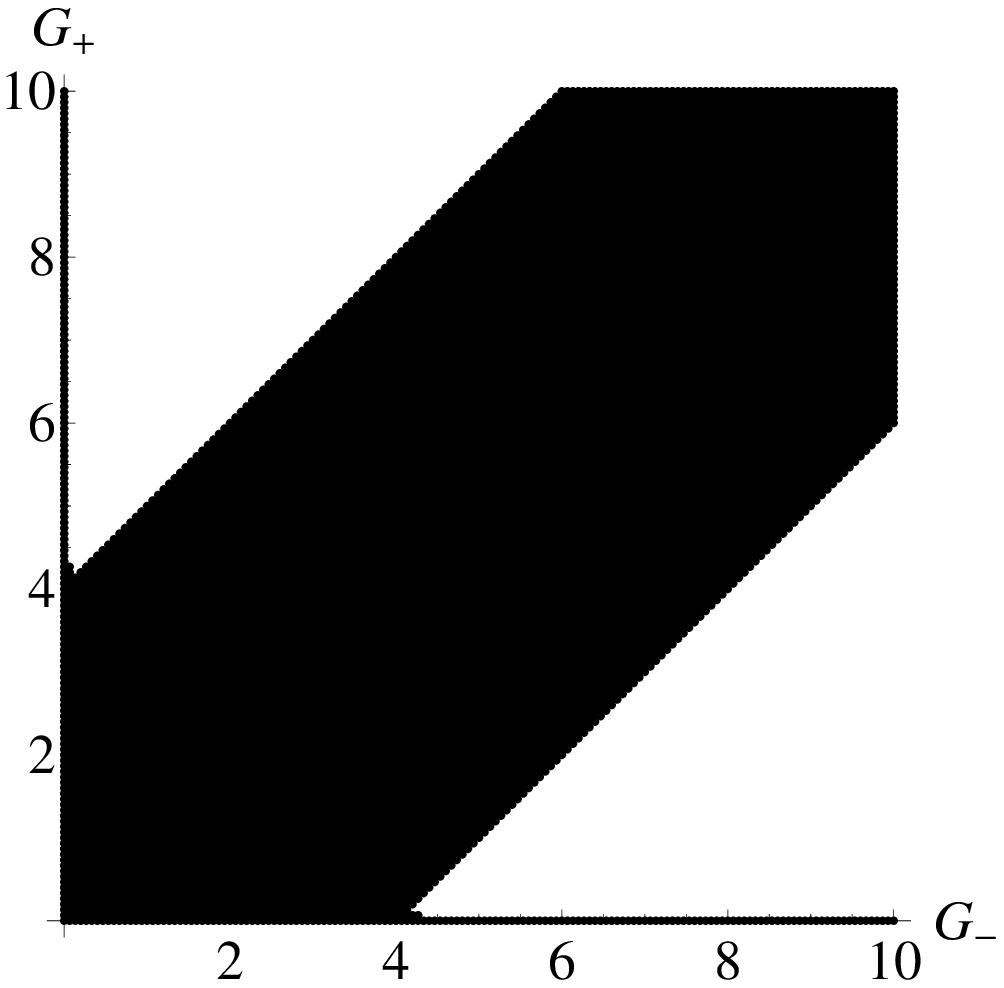} }\\ 
        \subfloat[]{  \label{fig:third}  \includegraphics[width=0.4\textwidth]{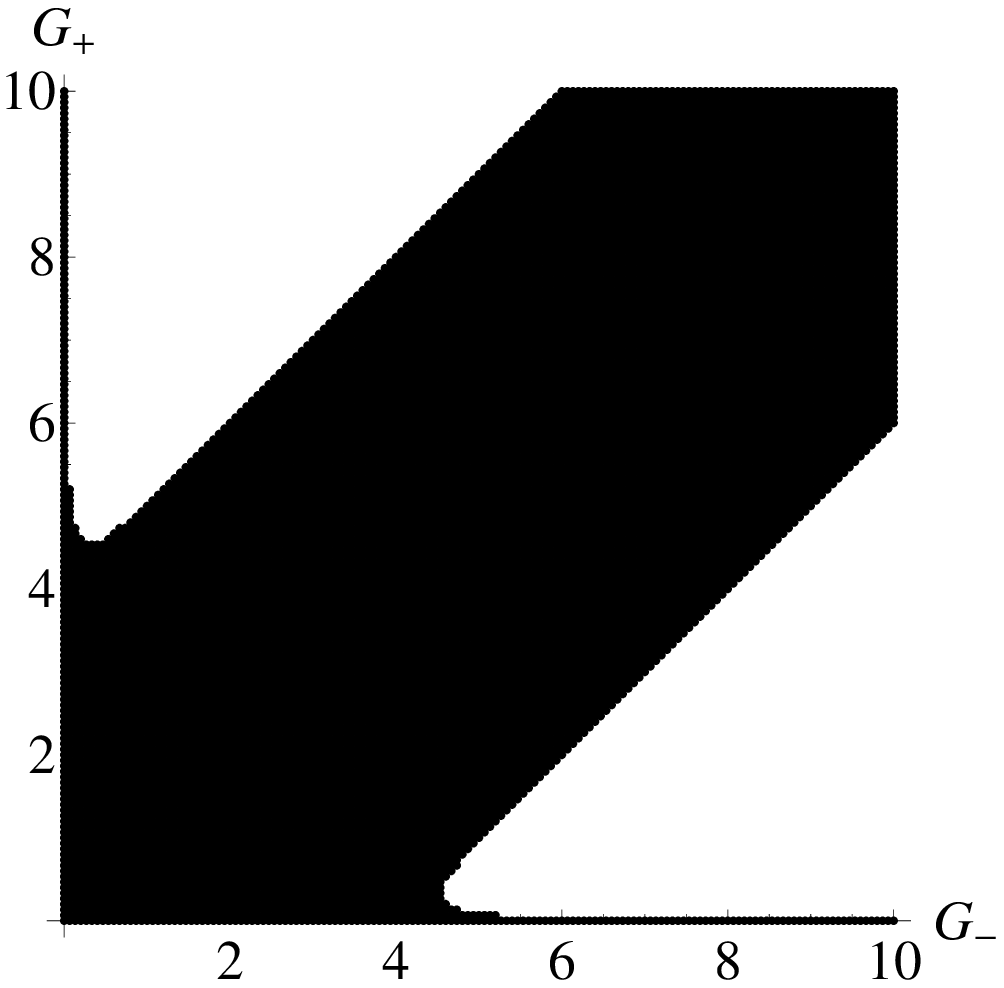} }
        \subfloat[]{  \label{fig:fourth} \includegraphics[width=0.4\textwidth]{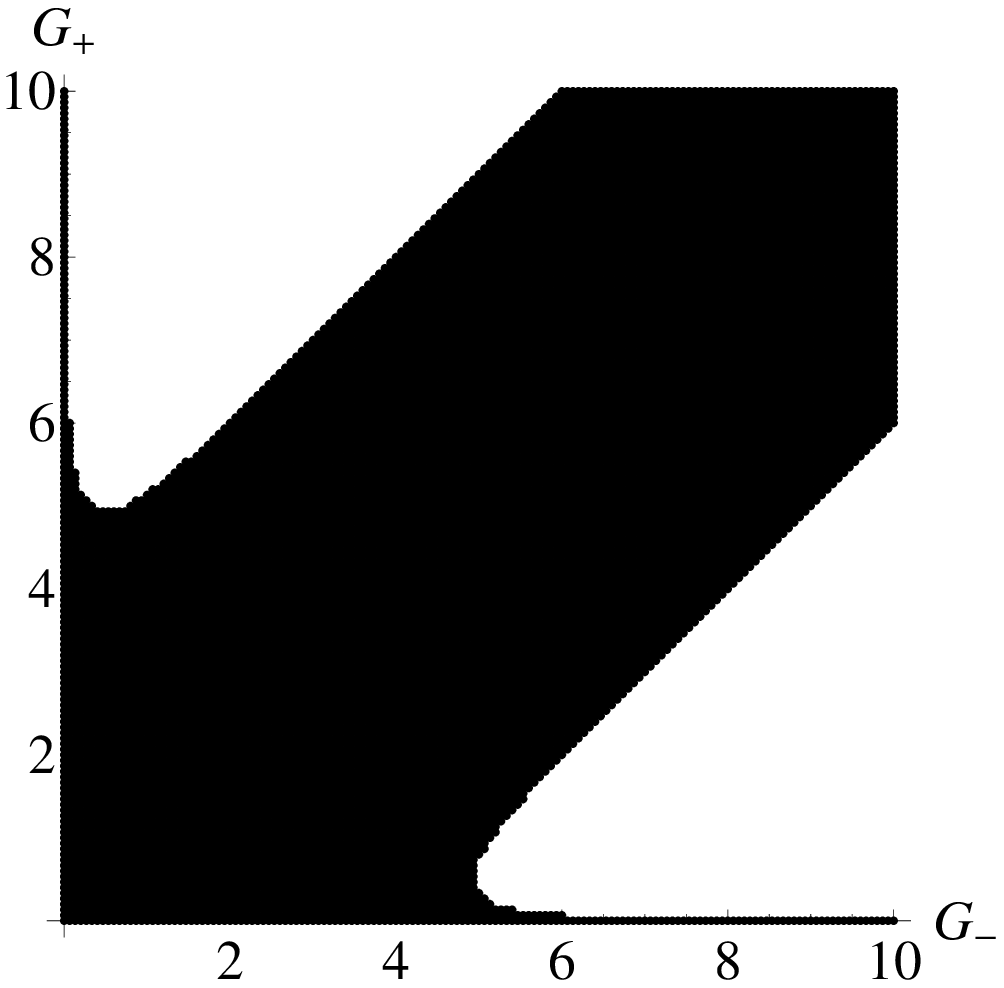} }
    \end{center}
    \caption{
          The shaded region depicts the values of the coupling parameters $0 \leq G_+,\,G_-\leq 10$ for which (\ref{bcs_eqn},\ref{gees},\ref{pf}) has real spectrum for the following choices: (a) $\delta=2$, $L=8$, $m=6$, and $M=3$; (b) $\delta=2$, $L=8$, $m=4$, and $M=2$; (c) $\delta=2$, $L=8$, $m=6$, and $M=2$, with blocked levels at $\epsilon_2=-5$ and $\epsilon_6=3$; (d) $\delta=2$, $L=8$, $m=6$, and $M=2$, with blocked levels at $\epsilon_4=-1$ and $\epsilon_7=5$.}
\end{figure}

Below we will instead consider the case for which  $G_+$ and $G_-$ are both real-valued. Although this does not lead to a Hermitian Hamiltonian, nonetheless the spectrum is real-valued in some region of the coupling paramater space.
Throughout we will work with a {\it picket fence} model whereby the $\epsilon_j$ are uniformly and symmetrically distributed around zero. In particular we write 
\begin{align}
\epsilon_j = \left(j-\frac{L+1}{2}\right)\delta.
\label{pf}
\end{align}
where the level spacing $\delta$ provides an energy scale for the system. Through numerical diagonalisation of the Hamiltonian it is found that there are clearly identifiable regions where the spectrum is real. Illustrative cases are depicted in Fig. 1. It is evident that the boundary lines separating real-valued and complex-valued spectra are approximately
\begin{align}
G_+-G_-=\pm 2\delta  
\label{lines}
\end{align}     
for sufficiently large $G_+$ and $G_-$.

On the lines of the coupling parameter space given by (\ref{lines}) we will describe a class of free quasi-particle excitations exhibiting generalised exclusion statistics. Our analysis hereafter will be conducted using exact results. In order to present the exact Bethe ansatz solution for the Hamiltonian it is useful to make a change of variable. We parameterise the coupling constants $G_\pm$ through variables $\alpha$, $\eta$ such that 
\begin{align}
\alpha&=\frac{1}{2}\ln \left(\frac{G_+}{G_-}\right), \qquad\quad
\eta= \frac{1}{2}\left(G_+ - G_-\right).  
\label{change}
\end{align}   
Inverting these relations gives
\begin{align*}
G_+&=\frac{2\eta e^{\alpha}}{e^\alpha - e^{-\alpha}},\qquad \quad
G_-=\frac{2\eta e^{-\alpha}}{e^\alpha - e^{-\alpha}}.
\end{align*} 
To simplify the discussion we will restrict attention to the subspace of unblocked states. The Hamiltonain is block-diagonal on this subspace with sectors determined by the eigenvalues $M$ of the pair number operator $N$. For each $M$ the dimension of the subspace within the space of unblocked states is $L!/(M!(L-M)!)$. The total space of unblocked states has dimension $2^L$, with the exact solution for the energy spectrum given by  
\begin{align}
E=2\sum_{j=1}^M v_j
\label{nrg}
\end{align}
where the $v_j$ are solutions of the Bethe ansatz equations \cite{dl04}
\begin{align}
e^{2\alpha} \prod_{l=1}^L(v_k-\epsilon_l+\eta/2) \prod^M_{j\neq k}(v_k-v_j-\eta) 
&= \prod_{l=1}^L(v_k-\epsilon_l-\eta/2) \prod^M_{j\neq k}(v_k-v_j+\eta).    
\label{bae1}
\end{align}
Before proceeding to an analysis of excitations characterised by the exact solution, we first need to recall and develop  
some results concerning the algebraic structure underlying the exact solution of the model.

\section{The $gl(2)$-Yangian, Quantum Inverse Scattering Method, and the algebraic Bethe ansatz}

The $gl(2)$-Yangian, which is denoted $Y[gl(2)]$, is an infinite-dimensional algebra with generators $\{T^j_k[l]\,:\,j,k=1,2;\,l=1,...,\infty\}$. In order to specify the algebraic relations, it is customary to use generating functions dependent on a complex variable \cite{cp91}
\begin{align}
T^j_k(u)=\sum_{l=0}^\infty T^j_k[l]u^{-l}, \qquad u\in\,{\mathbb C}
\label{gen}
\end{align}
and then impose the commutation relations
\begin{align}
(u-v)[T^j_l(u),\,T^k_m(v)]=\eta (T^k_l(v)T^j_m(u)-T^k_l(u)T^j_m(v)).
\label{ycomm}
\end{align} 
For later use we note that the algebra admits a homomorphism $\phi: Y[gl(2)] \rightarrow Y[gl(2)]$
\begin{align}
\phi(T^j_k(u))=\rho(u)T^j_k(u) 
\label{yhom}
\end{align}
where $\rho(u)$ is any function of $u$. 

Each irreducible, finite-dimensional, highest-weight $Y[gl(2)]$-module is characterised by a highest-weight vector $|\Psi\rangle$ such that \cite{cp91} 
\begin{align*}
T^1_1(u)|\Psi\rangle &= a(u) |\Psi\rangle , \\
T^2_2(u)|\Psi\rangle &= d(u) |\Psi\rangle , \\
T^1_2(u)|\Psi\rangle &= 0
\end{align*} 
where $a(u)$ and $d(u)$ are monic polynomials. 
Given such a module, we may obtain an equivalent module-action through (\ref{yhom}) with $\hat{a}(u)=\rho(u)a(u),\,\hat{d}(u)=\rho(u)d(u)$. The condition on equivalences of module-actions is expressed through the {\it Drinfeld polynomial} $P(u)$ \cite{cp91} for which 
\begin{align}
\frac{P(u-\eta)}{P(u)}=\frac{a(u)}{d(u)}. 
\label{hw}
\end{align}
It is easily verified that 
$$\frac{a(u)}{d(u)}=\frac{\hat{a}(u)}{\hat{d}(u)},$$
signifying the equivalence of the actions.

Central elements are generated by the quantum determinant 
\begin{align*}
{\mathcal D}(u)&=T^1_1(u)T^2_2(u+\eta)-T^2_1(u)T^1_2(u+\eta)   \\
&= T^2_2(u+\eta)T^1_1(u)-T^2_1(u+\eta)T^1_2(u) 
\end{align*}
which commutes with the generating functions (\ref{gen}).
The quantum determinant ${\mathcal D}(u)$ takes the eigenvalue
$\mu(u)= a(u)d(u+\eta)$ on each irreducible, finite-dimensional, highest-weight module. In contrast to the Drinfeld polynomial the quantum determinant can distinguish equivalent module-actions, but might not distinguish inequivalent ones. For example consider a one-dimensional module with $a(u)=d(u)=1$. Using the homomorphism (\ref{yhom}) we then have for $\hat{a}(u)=\hat{d}(u)=\rho(u)$ 
\begin{align*}  
P(u)&=1, \\
\mu(u)&=\rho(u)\rho(u+1).
\end{align*}
Alternatively consider a highest-weight vector $|\Psi\rangle$ with the property that there exists $v\in\,{\mathbb C}$ such that $a(v)=d(v)=0$. It follows from the algebraic relations (\ref{ycomm}) that 
\begin{align}
|\tilde{\Psi}\rangle = T^2_1(v) |\Psi\rangle 
\label{newpsi}
\end{align}
satisfies 
\begin{align*}
T^1_1(u)|\tilde{\Psi}\rangle &= \tilde{a}(u) |\tilde{\Psi}\rangle , \\
T^2_2(u)|\tilde{\Psi}\rangle &= \tilde{d}(u) |\tilde{\Psi}\rangle , \\
T^1_2(u)|\tilde{\Psi}\rangle &= 0
\end{align*}  
with 
\begin{align}
\tilde{a}(u)&=\frac{u-v+\eta}{u-v}a(u), \label{newa}\\
\tilde{d}(u)&=\frac{u-v-\eta}{u-v}d(u),  \label{newd}
\end{align}
i.e.  $|\tilde{\Psi}\rangle$ satisfies the conditions of a highest-weight vector. Here, either $|\tilde{\Psi}\rangle$ vanishes which is necessarily the case if the corresponding module is irreducible, or the module with highest-weight vector $|\Psi\rangle$ is reducible. It is straightforward to check that  $|{\Psi}\rangle$ and $|\tilde{\Psi}\rangle$ admit the same eigenvalue  for the quantum determinant ${\mathcal D}(u)$, 
 $$ \mu(u)=a(u)d(u+\eta)=\tilde{a}(u)\tilde{d}(u+\eta).$$

The fundamental two-dimensional $Y[gl(2)]$-module $V(u)$ with basis $\{|1\rangle,\,|2\rangle\}$ admits the action
\begin{align*}
T^1_1(u)|1\rangle&= (u-\eta/2)|1\rangle, &  T^1_1(u)|2\rangle&= (u+\eta/2)|2\rangle , \\
T^1_2(u)|1\rangle&= 0, &  T^1_2(u)|2\rangle&= \eta|1\rangle , \\
T^2_1(u)|1\rangle&= \eta|2\rangle, &  T^2_1(u)|2\rangle&= 0 , \\
T^2_2(u)|1\rangle&= (u+\eta/2)|1\rangle, &  T^2_2(u)|2\rangle&= (u-\eta/2)|2\rangle.
\end{align*}  
Since $Y[gl(2)]$ is a {\it bialgebra} there is a co-product $\Delta: Y[gl(2)]\rightarrow Y[gl(2)] \otimes Y[gl(2)]$ given by   
\begin{align}
\Delta(T^j_k(u))= \sum_{l=1}^2 T^j_l(u) \otimes T^l_k(u-\epsilon)
\label{cp}
\end{align} 
for arbitrary $\epsilon \in\,{\mathbb C}$. 
Iterating the co-product action permits the the construction of tensor product modules
$$V(u;\epsilon_1,...\epsilon_L)=V(u-\epsilon_1)\otimes .... \otimes V(u-\epsilon_L) $$
with highest weight vector
\begin{align}
|\Psi\rangle=|1\rangle^{\otimes L}. 
\label{hwstate}
\end{align}
For generic choices of the $\epsilon_j$ this module is irreducible with Drinfeld polynomial \cite{cp91}
\begin{align}
P(u)=\prod_{j=1}^L(u-\epsilon_j+\eta/2).
\label{drinpoly}
\end{align}
 
By using $Y[gl(2)]$-modules, abstract integrable models can be formulated as follows in the framework of the Quantum Inverse Scattering Method \cite{ks79,tf79}. For $\alpha\in\, {\mathbb C}$ define a {\it transfer matrix} as 
$$t(u)=e^{-\alpha}T^1_1(u)+e^\alpha T^2_2(u). $$
From the commutation relations (\ref{ycomm}) it follows that 
$$\left[t(u),\,t(v)\right]=0 \qquad \quad \forall u,\,v\in\,{\mathbb C} $$
Through the above relation the transfer matrix is a generator of conserved (i.e. mutually commuting) operators for an abstract quantum system. A Hamiltonian can be defined as a polynomial function of the conserved operators. For the specific case of the Hamiltonian (\ref{bcs_eqn}) subject to (\ref{gees}) we refer to \cite{dl04}.

The algebraic Bethe ansatz provides a means to diagonalise the transfer matrix on a highest-weight module with highest weight vector $|\Psi\rangle$. The eigenvectors of the transfer matrix are taken to be of the form
$$|v_1,v_2,...,v_M\rangle = \prod_{j=1}^M T^2_1(v_j)|\Psi\rangle$$ 
where the ordering in the product is inconsequential, since the $T^2_1(v_j)$ commute. From the commutation relations (\ref{ycomm}) it is found that 
\begin{align}
t(u)|v_1,v_2,...,v_M\rangle &= \lambda(u)|v_1,v_2,...,v_M\rangle \nonumber \\
&\quad -\sum_{k=1}^M \frac{e^{-\alpha}\eta\,a(v_k)}{u-v_k}\left(\prod_{j\neq k}^M \frac{v_k-v_j+\eta}{v_k-v_j}\right)T^2_1(u)|v_1,....,v_{j-1},v_{j+1}...,v_M\rangle \nonumber \\
&\quad +\sum_{k=1}^M \frac{e^\alpha \eta\,d(v_k)}{u-v_k}\left(\prod_{j\neq k}^M \frac{v_k-v_j-\eta}{v_k-v_j}\right)T^2_1(u)|v_1,....,v_{j-1},v_{j+1}...,v_M\rangle
\label{action}
\end{align}
with
\begin{align*}
\lambda(u)= e^{-\alpha}a(u)\prod_{j=1}^M\frac{u-v_j+\eta}{u-v_j}+e^\alpha d(u)\frac{u-v_j-\eta}{u-v_j}.
\end{align*}
Thus, if for all $k=1,...,M$, 
\begin{align}
a(v_k)\prod_{j\neq k}^M \left(v_k-v_j+\eta\right)
=e^{2\alpha} d(v_k)\prod_{j\neq k}^M \left(v_k-v_j-\eta\right),
\label{bae2}
\end{align} 
the {\it unwanted} terms of the form $|v_1,....,v_{j-1},v_{j+1}...,v_M\rangle$ cancel. When this occurs either $|v_1,v_2,...,v_M\rangle$ is an eigenstate of the transfer matrix, or $|v_1,v_2,...,v_M\rangle$ vanishes. In the case that the state vanishes we refer to the corresponding solution set $\{v_1,v_2,...,v_M\}$ as {\it spurious}. In view of the homomorphism (\ref{yhom}), we can minimise the occurrence of spurious solutions sets in some instances by an appropriate choice of $\rho(u)$ to express (\ref{bae2}) in the form
\begin{align}
P(v_k-\eta)\prod_{j\neq k}^M \left(v_k-v_j+\eta\right)
=e^{2\alpha} P(v_k)\prod_{j\neq k}^M \left(v_k-v_j-\eta\right).
\label{bae3}
\end{align} 
For the choice of highest weight state given by (\ref{hwstate}) we obtain 
\begin{align*}
a(u)&=\prod_{l=1}^L(u-\epsilon_l-\eta/2),     \\
d(u)&=\prod_{l=1}^L(u-\epsilon_l+\eta/2)
\end{align*}
and find that  (\ref{hw}) holds for (\ref{drinpoly}). Substituting the above expressions into (\ref{bae2}) leads to (\ref{bae1}).

Our main objective in the remainder of this work is to determine, for the specific case of the Hamiltonian (\ref{bcs_eqn}) subject to (\ref{gees},\ref{pf}) and restricted to the lines (\ref{lines}), those instances where
$|v_1,v_2,...,v_M\rangle$ is an eigenstate and those instances where $|v_1,v_2,...,v_M\rangle$ vanishes. Through this investigation it will be seen how a class of free quasi-particle excitations exhibiting generalised exclusion statistics is uncovered.  

\section{Generalised exclusion statistics}

Restricting to the lines (\ref{lines}), which is equivalent to setting $\delta=\eta$, 
the Bethe ansatz equations (\ref{bae1}) are expressible as
\begin{align}
& \prod_{l=1}^{L-1}\left(v_k-\eta\left(l-\frac{L}{2}\right)\right) \nonumber \\
&\qquad \times \left(e^{2\alpha}\left(v_k+\frac{\eta L}{2}\right)\prod^M_{j\neq k}(v_k-v_j-\eta)-\left(v_k-\frac{\eta L}{2}\right)\prod^M_{j\neq k}(v_k-v_j+\eta)\right)=0 
\label{specbae}
\end{align}
for each $k=1,...,M$. It is instructive to first consider the cases $M=1$ and $M=2$ before addressing the general case.  
\subsection{The one Cooper pair sector $M=1$}

In this sector the dimension of the Hilbert space is $L$. Setting $M=1$ and $v_1\equiv v$ in (\ref{specbae}) leads to 
\begin{align*}
\left(e^{2\alpha}\left(v+\frac{\eta L}{2}\right)-\left(v-\frac{\eta L}{2}\right)\right)\prod_{l=1}^{L-1}\left(v-\eta\left(l-\frac{L}{2}\right)\right)=0. 
\end{align*}
The above is a polynomial equation of order $L$, the roots of which can be stated explicitly. There is one root which is $\alpha$-dependent, viz.
$$v=-\frac{\eta L}{2}\left(\frac{e^\alpha+e^{-\alpha}}{e^\alpha-e^{-\alpha}}\right),$$
while the remaining $\alpha$-independent roots are elements from the set 
\begin{align}
{\mathcal S}= \{\delta(k-L/2):k=1,...,L-1 \}. 
\label{s}
\end{align} 

To check whether any of these roots is spurious we consider the following action which is obtained through iterated use of the co-product (\ref{cp}): 
\begin{align*}
T^2_1(u)|\Psi\rangle &= \eta\sum_{j=1}^L \prod_{k=1}^{j-1} (u-\epsilon_k-\eta/2) \prod_{l=j+1}^L(u-\epsilon_l+\eta/2) S_j^+|\Psi\rangle \\
&= \eta\prod_{k=1}^{L-1}(u-\delta(k-L/2))\sum_{j=1}^L S_j^+ |\Psi\rangle. 
\end{align*}
For the $\alpha$-dependent root, substitution into the above expression yield an eigenstate of the Hamiltonian. However,
at first sight it appears that the roots from ${\mathcal S}$ are spurious. The state $T^2_1(v)|\Psi\rangle$ vanishes for $v\in\,{\mathcal S}$ due to the co-efficient polynomial $\prod_{k=1}^{L-1}(v-\delta(k-L/2))$. Rescaling the state by this polynomial would then lead to the {\it same} eigenstate $\sum_{j=1}^LS_j^+|\Psi\rangle$ for {\it all} $v\in\,{\mathcal S}$. However directly applying the Hamilonian to this state confirms that it is not an eigenstate. The problem with this approach is that rescaling by $\prod_{k=1}^{L-1}(v-\delta(k-L/2))$ means the unwanted terms in (\ref{action}) no longer cancel. 

Nonetheless we can confirm that the roots from ${\mathcal S}$ are not spurious. Starting with a generic value for $\eta$ we have 
\begin{align*}
T^2_1(u)|\Psi\rangle &= \eta\sum_{j=1}^L \prod_{k=1}^{j-1} (u-\epsilon_k-\eta/2) \prod_{l=j+1}^L(u-\epsilon_l+\eta/2) S_j^+|\Psi\rangle \\
&=\eta\prod_{\beta=1}^{\mu-1}(u-\epsilon_\beta-\eta/2)\prod_{\gamma=\mu+1}^L(u-\epsilon_\gamma+\eta/2) \\
&\quad \times \sum_{j=1}^L \frac{\prod_{k=1}^{j-1} (u-\epsilon_k-\eta/2)}{\prod_{\gamma=\mu+1}^L(u-\epsilon_\gamma+\eta/2)} \frac{\prod_{l=j+1}^L(u-\epsilon_l+\eta/2)}{\prod_{\beta=1}^{\mu-1}(u-\epsilon_\beta-\eta/2)} S_j^+|\Psi\rangle .
\end{align*}
For $\mu=1,...,L-1$ let 
$$\Upsilon_\mu(u)=
{\eta}{\prod_{\beta=1}^{\mu-1}(u-\epsilon_\beta-\eta/2)\prod_{\gamma=\mu+1}^L(u-\epsilon_\gamma+\eta/2)}. $$
Then we define rescaled states
\begin{align*}
|\Phi_{\mu}(u)\rangle &= \frac{1}{\Upsilon_\mu(u)}T^2_1(u)|\Psi\rangle \\
&=\sum_{j=1}^{\mu-1} \frac{\prod_{k=1}^{j-1} (u-\epsilon_k-\eta/2)}{\prod_{\gamma=\mu+1}^L(u-\epsilon_\gamma+\eta/2)} \frac{\prod_{l=j+1}^L(u-\epsilon_l+\eta/2)}{\prod_{\beta=1}^{\mu-1}(u-\epsilon_\beta-\eta/2)} S_j^+|\Psi\rangle  +S^+_\mu |\Psi\rangle \\
&\qquad+\sum_{j=\mu+1}^L \frac{\prod_{k=1}^{j-1} (u-\epsilon_k-\eta/2)}{\prod_{\gamma=\mu+1}^L(u-\epsilon_\gamma+\eta/2)} \frac{\prod_{l=j+1}^L(u-\epsilon_l+\eta/2)}{\prod_{\beta=1}^{\mu-1}(u-\epsilon_\beta-\eta/2)} S_j^+|\Psi\rangle \\
&=\sum_{j=1}^{\mu-1}  \prod_{\beta=j}^{\mu-1}\frac{(u-\epsilon_{\beta}-\delta+\eta/2)}{(u-\epsilon_\beta-\eta/2)} S_j^+|\Psi\rangle  +S^+_\mu |\Psi\rangle \\
&\qquad+\sum_{j=\mu+1}^L \frac{\prod_{k=1}^{j-1} (u-\epsilon_k-\eta/2)}{\prod_{\gamma=\mu+1}^L(u-\epsilon_\gamma+\eta/2)} \frac{\prod_{l=j+1}^L(u-\epsilon_l+\eta/2)}{\prod_{\beta=1}^{\mu-1}(u-\epsilon_\beta-\eta/2)} S_j^+|\Psi\rangle .
\end{align*}
The Bethe ansatz equations (\ref{bae1}) for $M=1$ may be written as
\begin{align*}
\frac{\prod_{k=1}^{j-1} (u-\epsilon_k-\eta/2)}{\prod_{\gamma=\mu+1}^L(u-\epsilon_\gamma+\eta/2)}
&= e^{2\alpha} \frac{\prod_{\gamma=1}^\mu (u-\epsilon_\gamma+\eta/2)}{\prod_{k=j}^L(u-\epsilon_k-\eta/2)}. 
\end{align*}
This then leads to 
\begin{align*}
|\Phi_{\mu}(u)\rangle
&=\sum_{j=1}^{\mu-1}  \prod_{\beta=j}^{\mu-1}\frac{(u-\epsilon_{\beta}-\delta+\eta/2)}{(u-\epsilon_\beta-\eta/2)} S_j^+|\Psi\rangle  +S^+_\mu |\Psi\rangle \\
&\qquad+e^{2\alpha}\sum_{j=\mu+1}^L  \frac{u-\epsilon_\mu+\eta/2}{u-\epsilon_j-\eta/2}
\prod_{\beta=1}^{\mu-1}\frac{u-\epsilon_\beta+\eta/2}{u-\epsilon_{\beta+1}+\eta/2}
\prod_{l=j+1}^L \frac{u-\epsilon_{l-1}-\eta/2}{u-\epsilon_l-\eta/2} S_j^+|\Psi\rangle .
\end{align*}
Next let $\eta=\delta$ giving
\begin{align}
|\Phi_{\mu}(u)\rangle&=\sum_{j=1}^{\mu}   S_j^+|\Psi\rangle  +e^{2\alpha}\sum_{j=\mu+1}^L  \frac{u-\epsilon_{1}+\delta/2}{u-\epsilon_L-\delta/2} S_j^+|\Psi\rangle, \nonumber \\
|\Phi_{\mu}(\delta(\mu-L/2))\rangle&=\sum_{j=1}^{\mu}   S_j^+|\Psi\rangle  +e^{2\alpha}\sum_{j=\mu+1}^L  \frac{(\mu-L/2)\delta-(1-L)\delta/2+\delta/2}{(\mu-L/2)\delta-(L-1)\delta/2-\delta/2} S_j^+|\Psi\rangle \nonumber\\
&=\sum_{j=1}^{\mu}   S_j^+|\Psi\rangle  +\frac{e^{2\alpha}\mu}{\mu-L}\sum_{j=\mu+1}^L   S_j^+|\Psi\rangle .
\label{newhwstates}
\end{align}
It can be verified by direct calculation that the expressions for $|\Phi_{\mu}(\delta(\mu-L/2))\rangle,\,\mu=1,...,L-1$ as given by (\ref{newhwstates}) are eigenstates of the Hamiltonian (\ref{bcs_eqn}) subject to (\ref{gees},\ref{pf},\ref{lines}). Along with the eigenstate associated with the $\alpha$-dependent root, this provides a complete set of $L$ eigenstates for the sector $M=1$.
    
\subsection{The two Cooper pair sector $M=2$}

In this sector the dimension of the Hilbert space is $L(L-1)/2$. The first important observation to make from the previous Subsection is that when $\eta=\delta$ the roots of the Bethe equations can be clearly demarcated into two groups, those which are $\alpha$-independent and those which are not. This provides three sub-cases to consider.

\subsubsection{Two $\alpha$-independent roots.} \label{tworoots}
From the Bethe ansatz equations (\ref{specbae}) it is seen that a solution set is {\it formally} obtained by choosing any two elements from the set ${\mathcal S}$ given by Eq. (\ref{s}), including the case when the roots are equal. In this respect the excitations have the character of two free quasi-particles. However some solution sets are spurious. To identify those instances, we first note that the one Cooper pair sector eigenstates (\ref{newhwstates}) are $Y[gl(2)]$ highest weight states of the form $(\ref{newpsi})$ with suitable rescaling. Specifically, choosing $v_1=\delta(\mu-L/2)$ leads us to consider 
\begin{align*}
T^1_1(u)|\Phi_\mu (\delta(\mu-L/2))\rangle &= \tilde{a}(u)|\Phi_\mu (\delta(\mu-L/2))\rangle    , \\
T^2_2(u)|\Phi_\mu (\delta(\mu-L/2))\rangle &= \tilde{d}(u)|\Phi_\mu (\delta(\mu-L/2))\rangle 
\end{align*}   
with
\begin{align*}
\tilde{a}(u)&=(u-\delta(\mu-1-L/2))\prod_{l\neq \mu}^L(u-\delta(l-L/2))   , \\
\tilde{d}(u)&=(u-\delta(\mu+1-L/2))\prod_{l\neq \mu+1}^L(u-\delta(l-1-L/2))     
\end{align*}
and the associated Drinfeld polynomial 
\begin{align*}
P(u)=\prod_{l\neq \mu,\mu+1}^L(u-\delta(l-1-L/2)) 
\end{align*}  
such that (\ref{hw}) holds. For this Drinfeld polynomial we find that the set of $\alpha$-independent roots for the Bethe ansatz equations (\ref{bae3}), from which we can choose $v_2$, is the restricted set
\begin{align*}
{\mathcal S}'= \{\delta(k-L/2):k=1,...,L-1; \quad k\neq \mu-1, \mu, \mu+1  \}. 
\end{align*} 
Here we observe the manifestation of generalised exclusion statistics. Having first chosen the Bethe root 
$v_1=\delta(\mu-L/2)$ we find that not only can we not choose it again (as is the case for the familiar fermionic exclusion principle), we also cannot choose the ``neighbouring'' roots $\delta(\mu\pm 1-L/2)$. Bearing this in mind, a simple counting argument shows that the number of eigenstates in the $M=2$ sector where both roots are $\alpha$-independent is given by 
$(L-2)(L-3)/2$.

\subsubsection{One $\alpha$-independent root.}
Choosing $v_1\in\,{\mathcal S}$ 
the Bethe ansatz equation for $v_2\notin\,{\mathcal S}$ is the quadratic equation 
\begin{align*}
e^{2\alpha} (v_2+\delta L/2)(v_2-v_1-\delta)&=(v_2-\delta L/2)(v_2-v_1+\delta).
\end{align*}
The solution reads 
%
%
$$v_2=\frac{2(1-e^{2\alpha}) v_1 +\delta (1+e^{2\alpha})(L-2)\pm\sqrt{D}}{4(1-e^{2\alpha})}$$
where 
\begin{align*}
D&=\delta^2(e^{2\alpha}+1)^2(L-2)^2+4(e^{2\alpha}-1)^2(v_1^2+2\delta^2 L) 
+4\delta(e^{4\alpha}-1)(L+2) v_1. 
\end{align*}

While there are generally two solutions for $v_2$, two special cases need to be re-examined in closer detail.
Choosing $v_1=\delta(1-L/2)$ the equation for $v_2$ is 
\begin{align*}
e^{2\alpha} (v_2+\delta L/2)(v_2-v_1-\delta)&=(v_2-\delta L/2)(v_2-v_1+\delta) \\
e^{2\alpha}(v_2+\delta L/2)(v_2-2\delta +\delta L/2)&= (v_2-\delta L/2)(v_2+\delta L/2). 
\end{align*}
Formally there are two solutions for $v_2$, but the case $v_2=-\delta L/2$ is $\alpha$-independent and spurious by the reasoning presented in the preceeding Subsubsection. The $\alpha$-dependent  root is 
$$v_2=\frac{(L/2+(L/2-2)e^{2\alpha})\delta}{1-e^{2\alpha}}. $$
Similarly, choosing $v_1=\delta(L/2-1)$ the equation for $v_2$ is 
\begin{align*}
e^{2\alpha} (v_2+\delta L/2)(v_2-v_1-\delta)&=(v_2-\delta L/2)(v_2-v_1+\delta) \\
e^{2\alpha}(v_2+\delta L/2)(v_2-\delta L/2)&= (v_2-\delta L/2)(v_2-\delta L/2+2\delta).
\end{align*}
In this instance the $\alpha$-independent solution $v_2=\delta L/2$ is spurious. The $\alpha$-dependent  root is 
$$v_2=\frac{((L/2-2)+e^{2\alpha}L/2)\delta}{1-e^{2\alpha}}. $$
These results indicate that the number of eigenstates in the $M=2$ sector, where only one root is $\alpha$-independent, is given by $2(L-2)$.   



\subsubsection{No $\alpha$-independent roots.}
Since the the dimension of the Hilbert space for $M=2$ is $L(L-1)/2$, and we have accounted for $(L-2)(L-3)/2$ solutions for two $\alpha$-independent roots and $2(L-2)$ solutions for one $\alpha$-independent roots, there can only be one non-spurious solution with both roots being $\alpha$-dependent. Although we have not been able to derive this result in the sense of producing an explicit general formula for the solution, we have checked the low-dimensional cases $L=4,5,6$ with the results tabulated in Appendix A. Specifically, we have verified that for these cases our analyses above are in complete agreement with results obtained by direct diagonalisation of the Hamiltonian to obtain the eigenspectrum. In particular, direct diagonalisation confirms the picture that the $\alpha$-independent roots are associated with free quasi-particle excitations with generalised exclusion statistics.

\subsection{The case of general $M$}

For a given $L$ and $M$ let $P$, with $P\leq M$, denote the number of $\alpha$-independent roots within a set of $M$ roots. 
Further let $n(L,M,P)$ denote the number of non-spurious solutions for each set of these quantities. From our previous discussions we have $n(L,0,0)=1$, $n(L,1,1)=L-1$, $n(L,1,0)=1$, $n(L,2,2)=(L-2)(L-3)/2$, $n(L,2,1)=2(L-2)$ and $n(L,2,0)=1$.

Next we will prove a formula for $n(L,M,M)$, i.e., the case when all roots are $\alpha$-independent. We saw in the previous Subsection for $M=2$ that if we choose a particular root $v_1=\delta(\mu-L/2)$, the generalised exclusion principle prohibits the choice $v_2=\delta(\mu\pm1-L/2)$ to obtain an eigenstate. This result generalises for arbitrary $M$ in a straightforward manner, viz. given a solution set for the $M-1$ Cooper pair sector 
$$\{v_j=\delta(\mu_j-L/2): j=1,...,M-1\}$$ 
we are prohibited from choosing any neighbouring roots $\delta(\mu_j\pm1-L/2)$ for $v_M$ to obtain an eigenstate in the $M$ Cooper pair sector. The proof is a matter of iterating the procedure described for $M=2$ in Subsubsection \ref{tworoots}. To determine $n(L,M,M)$ we need to count the number of $\alpha$-independent solution sets which respect to this exclusion rule.

Let 
$$\left[
\begin{array}{c} p \\ q \end{array} 
\right] $$
denotes the number of ways that $q$ identical objects can be placed in $p$ boxes such that the objects cannot be placed into adjacent boxes. This quantity satisfies the recursion relation 
$$\left[
\begin{array}{c} p \\ q \end{array} 
\right] =\left[
\begin{array}{c} p-1 \\ q \end{array} 
\right] + \left[
\begin{array}{c} p-2 \\ q-1 \end{array} 
\right]  $$
with the initial conditions 
\begin{align*}
\left[
\begin{array}{c} p \\ 0 \end{array} 
\right] &= 1 \qquad p\geq 0, \\
\left[
\begin{array}{c} 0 \\ q \end{array} 
\right] &=0\qquad q>0,  \\
\left[
\begin{array}{c} 1 \\ 1 \end{array} 
\right] &=1.  
\end{align*}
The recursion relation can be solved to obtain 
$$\left[
\begin{array}{c}
p \\ q \end{array}
\right]= \frac{(p+1-q)!}{(p+1-2q)!q!}. $$
The solution set ${\mathcal S}$ as given by (\ref{s}) contains $L-1$ elements, yielding  
$$n(L,M,M)=\left[
\begin{array}{c}
L-1 \\ M \end{array}
\right]= \frac{(L-M)!}{(L-2M)!M!}.$$

Although we have not been able to prove a general result for $n(L,M,P)$, we conjecture that it is given by 
\begin{align}
n(L,M,P)=\frac{(L-M)!M!}{(L-M-P)!(M-P)!(P!)^2}. 
\label{conj}
\end{align}
This formula agrees with the subcases $P=M$, and $M=0,1,2$ with $P\leq M$, discussed above. It also satisfies the required condition  
$$\sum_{P=0}^M n(L,M,P) = \frac{L!}{(L-M)!M!} $$ 
to account for all states in each sector of $M$ Cooper pairs. The above identity may be proved by considering binomial expansions of the terms in  
$$(1+x)^L=(1+x)^{L-M}(1+x)^M. $$

\section{Conclusion} The Hamiltonian (\ref{bcs_eqn}), subject to (\ref{gees},\ref{pf}),  admits an exact solution. On the subspace of unblocked states the energy eigenvalues are given by (\ref{nrg}) where the set $\{v_1,\,...\,v_M\}$ is a solution of the Bethe ansatz equations (\ref{bae1}). The energy expression is a simple sum of the Bethe roots, which leads to a natural interpretation of each Bethe root being associated with a quasi-particle. In this quasi-particle picture the states are generally bound rather than free, since a solution set for $M$ quasiparticles is not simply the union of one-body
solutions due to the coupled nature of (\ref{bae1}). In general the roots $v_j$ will depend on the two coupling parameters $G_+,\,G_-$, or equivalently, the variables $\alpha$ and $\eta$ as given by (\ref{change}).  

Setting $\eta=\delta$ corresponding to (\ref{lines}), where $\delta$ is the level spacing in (\ref{pf}), we find that the roots of the Bethe ansatz equations divide into two classes, viz, those which are $\alpha$-dependent and those which are $\alpha$-independent. From this perspective we made identify the lines (\ref{lines}) with deconfined excitations, those being associated to the $\alpha$-independent roots which do not occur for general $\eta$. From the form of the Bethe ansatz equations given by (\ref{specbae}), the $\alpha$-independent roots were determined explicilty. They are elements of the set ${\mathcal S}$ defined in (\ref{s}). The corresponding states were found to be highest weight states of the underlying $Y[gl(2)]$ algebraic structure. These roots are associated with free quasi-particles excitations, since a solution set for $M$ quasiparticles is a union of one-body solutions. However, a close examination of the states associated with these solutions led to the conclusion that there are spurious solution sets, which give rise to an interpretation of generalised exclusion statistics. We have conjectured the formula (\ref{conj}) for the number of non-spurious solution sets with $M$ roots, where $P$ of the roots are $\alpha$-independent. A proof of the result for $P=M$ was provided, and it was also found that (\ref{conj}) is valid in certain limiting cases.          

It was proposed in \cite{sbsvf04,svbsf04} that quantum criticality
may be identified by deconfined excitations at the critical point, which are not
found in phases adjacent to the critical point. It is interesting to note for the present study that the lines (\ref{lines}) on which the deconfined $\alpha$-independent excitations occur are in very close agreement with the boundary lines between the unbroken ${\mathcal PT}$-symmetric phase and a broken ${\mathcal PT}$-symmetric phase. The ${\mathcal PT}$-symmetry breaking phase boundary lines are clearly identified in Fig. 1, which was obtained by direct numerical diagonalisation of the Hamiltonian.  
 
\section*{Appendix A - Tables of roots of the Bethe ansatz equations}

In this Appendix we list the roots of the Bethe ansatz equations for some small sized systems $L=4,5,6$, $M=2$ and no blocked states, for when the coupling parameters correspond to the lines (\ref{lines}). The first column lists the energy eigenvalues obtained by direct diagonalisation, while the second column lists the roots of the Bethe ansatz equations (\ref{specbae}) associated to each energy eigenvalue through (\ref{nrg}). These results are consistent with conjecture (\ref{conj}) concerning the distribution of roots, and establish that the energy spectrum of the Hamiltonian on the lines (\ref{lines}) is real-valued for these cases.

\begin{table}
\begin{center} 
\begin{tabular}{|c|cc|}
\hline
  $E$& $(v_1,v_2)$&   \\ \hline & & \\[-5mm] \hline  & & 
 \\[-5mm]
  $0$ & $(\delta,-\delta)$ &
  \\[1mm] \hline
  & & \\[-5mm]
 $\dfrac{2 \left(1+e^{2 \alpha }\right) \delta }{1-e^{2 \alpha }}$ & 
$\left(\delta, \dfrac{2 e^{2 \alpha } \delta }{1-e^{2 \alpha }}\right)$,  $\left(-\delta, \dfrac{2 \delta }{1-e^{2 \alpha }}\right)$ & 
\\[1mm]\hline
& & \\[-5mm]
$\dfrac{(1 +e^{2 \alpha }  +\sqrt{9 -14 e^{2 \alpha } +9 e^{4 \alpha } })\delta}{1-e^{2 \alpha }}$ & $\left(0,\dfrac{(1 +e^{2 \alpha } +\sqrt{9-14 e^{2\alpha }+9 e^{4 \alpha }})\delta}{2(1-e^{2 \alpha })}\right) $ &
\\[1mm] \hline
 & & \\[-5mm]
$\dfrac{(1 +e^{2 \alpha }  -\sqrt{9 -14 e^{2 \alpha } +9 e^{4 \alpha } })\delta}{1-e^{2 \alpha }}$ & $\left(0,\dfrac{(1 +e^{2 \alpha } -\sqrt{9-14 e^{2\alpha }+9 e^{4 \alpha }})\delta}{2(1-e^{2 \alpha })}\right)$ &
\\[1mm] \hline
 & & \\[-5mm]
 \multirow{3}{*}{$\dfrac{6 \left(1+e^{2 \alpha }\right) \delta }{1-e^{2 \alpha }}$} & 
 $\left(\dfrac{\left(3+3 e^{2 \alpha }+\sqrt{1-14 e^{2 \alpha}+e^{4\alpha }}\right) \delta }{2 \left(1-e^{2 \alpha }\right)},\right.$& \\
 & $\quad\quad\left.\dfrac{\left(3+3 e^{2 \alpha } -\sqrt{1-14e^{2 \alpha}+e^{4 \alpha }}\right)\delta}{2\left(1-e^{2 \alpha}\right)}\right)$&
\\[4mm] \hline 
\end{tabular} 
\caption{The energy spectrum and associated solutions of the Bethe ansatz equations (\ref{specbae}) on the lines (\ref{lines}) for $L=4,\,M=2$. While the pairs $(\delta,0)$, $(0,-\delta)$ are solutions of (\ref{specbae}) they do not correspond to an eigenvalue of the Hamiltonian, giving rise to a generalised exclusion principle.}
\end{center}
 \end{table}

\begin{table}
\begin{center}
 \begin{tabular}{|c|cc|}
 \hline
  $E$& $(v_1,v_2)$&   \\ \hline & & \\[-5mm] \hline  & & 
 \\[-5mm]
 $2 \delta$ & $\left(\dfrac{3 \delta }{2},-\dfrac{\delta }{2}\right)$ &
 \\[1mm]\hline
  & & \\[-5mm]
 {$0$} & $\left(\dfrac{3 \delta }{2},-\dfrac{3 \delta }{2}\right)$ 
                       &
\\[1mm] \hline
& & \\[-5mm]
 $-2\delta$  & $\left(\dfrac{\delta }{2},-\dfrac{3 \delta }{2}\right)$ &
\\[1mm]\hline
 & & \\[-5mm]
$\dfrac{2 \left(2+e^{2 \alpha }\right) \delta }{1-e^{2 \alpha }}$ & $\left(\dfrac{3 \delta }{2}, \dfrac{(1 +5 e^{2 \alpha }) \delta }{2(1- e^{2 \alpha })}\right)$ & 
\\[1mm]\hline
 & & \\[-5mm]
$\dfrac{2 \left(1+2 e^{2 \alpha }\right) \delta }{1-e^{2 \alpha }}$ & $\left(-\dfrac{3 \delta }{2},\dfrac{\left(5+e^{2 \alpha }\right) \delta }{2 \left(1-e^{2 \alpha }\right)}\right)$  																																&
\\[1mm]\hline
 & & \\[-5mm]
$\dfrac{\left(3-\sqrt{9-16 e^{2 \alpha }+16e^{4\alpha} }\right) \delta }{1-e^{2 \alpha }}$& $\left(\dfrac{\delta }{2},\dfrac{\left(2+e^{2 \alpha }-\sqrt{9-16 e^{2 \alpha }+16 e^{4 \alpha }}\right) \delta }{2 \left(1-e^{2 \alpha }\right)}\right)$ &
\\[1mm]\hline
 & & \\[-5mm]
$\dfrac{\left(3+\sqrt{9-16e^{2\alpha}+16 e^{4 \alpha } }\right) \delta }{1-e^{2 \alpha }}$&$\left(\dfrac{\delta }{2},\dfrac{\left(2+e^{2 \alpha}+\sqrt{9-16 e^{2 \alpha }+16 e^{4 \alpha }}\right) \delta }{2 \left(1-e^{2 \alpha }\right)}\right)$&
\\[1mm] \hline
  & & \\[-5mm]
  $\dfrac{\left(3 e^{2 \alpha }-\sqrt{16-16 e^{2 \alpha }+9 e^{4 \alpha }}\right) \delta }{1-e^{2 \alpha }}$ & $\left(-\dfrac{\delta }{2},\dfrac{\left(1+2 e^{2 \alpha}-\sqrt{16-16 e^{2\alpha }+9 e^{4 \alpha }}\right) \delta }{2 \left(1-e^{2 \alpha }\right)}\right)$ &
 \\[1mm]\hline
 & & \\[-5mm]
 $\dfrac{\left(3 e^{2 \alpha }+\sqrt{16-16 e^{2 \alpha }+9 e^{4 \alpha }}\right) \delta }{1-e^{2 \alpha }}$ & $\left(-\dfrac{\delta }{2},\dfrac{(1 +2 e^{2 \alpha } +\sqrt{16-16e^{2 \alpha }+9 e^{4 \alpha }}) \delta }{2(1- e^{2 \alpha })}\right) $&
\\[1mm]\hline
 & & \\[-5mm]
 \multirow{3}{*}{$\dfrac{8 \left(1+e^{2 \alpha }\right) \delta }{1-e^{2 \alpha }}$} & 
 $
 \left(\dfrac{\left(4+4 e^{2 \alpha }+\sqrt{1-18 e^{2 \alpha }+e^{4 \alpha }}\right) \delta }{2(1-e^{2 \alpha })},\right.
 $& \\
 & $
 \quad\quad\left.\dfrac{\left(4+4 e^{2 \alpha }-\sqrt{1-18 e^{2 \alpha }+e^{4 \alpha }}\right) \delta}
 {2 (1-e^{2 \alpha})}\right)
 $& \\
 \hline
 \end{tabular}\\
 \caption{The energy spectrum and associated solutions of the Bethe ansatz equations (\ref{specbae}) on the lines (\ref{lines}) for $L=5,\,M=2$. While the pairs $(3\delta/2,\delta/2)$, $(\delta/2,-\delta/2)$, $(-\delta/2,-3\delta/2)$ are solutions of (\ref{specbae}) they do not correspond to an eigenvalue of the Hamiltonian, giving rise to a generalised exclusion principle.}
 \end{center}
 \end{table}

~~\\~~\\ 
\begin{table}
\begin{center}
\begin{tabular}{|c|cc|}
\hline
 $E$& $(v_1,v_2)$&   \\ \hline & & \\[-5mm] \hline  & & 
 \\[-5mm]
 $4 \delta $ & $(2\delta,0)$ &
\\[1mm] \hline
& & \\[-5mm] 
 $2 \delta $ & $(2\delta,-\delta)$ &
\\[1mm] \hline
& & \\[-5mm] 
 {$ 0 $} & $(\delta,-\delta),\,\,(2\delta,-2\delta)$ &
\\[1mm] \hline
& & \\[-5mm] 
 $-2 \delta $  & $(\delta,-2\delta)$ &
\\[1mm] \hline
& & \\[-5mm]
 $-4 \delta $ & $(0,-2\delta)$&
\\[1mm] \hline  
& & \\[-5mm] 
$\dfrac{\left(5-e^{2 \alpha }-\sqrt{9-18 e^{2 \alpha }+25 e^{4 \alpha }}\right) \delta }{1-e^{2 \alpha }}$& 
$\left(\delta,\dfrac{\left(3+e^{2 \alpha }-\sqrt{9-18 e^{2 \alpha }+25 e^{4 \alpha }}\right) \delta }{2 \left(1-e^{2 \alpha }\right)}\right)$ &
\\[4mm] \hline 
& & \\[-5mm]
$\dfrac{\left(5-e^{2 \alpha }+\sqrt{9-18 e^{2 \alpha }+25 e^{4 \alpha }}\right) \delta }{1-e^{2 \alpha }}$& $\left(\delta, \dfrac{\left(3+e^{2 \alpha }+\sqrt{9-18 e^{2 \alpha }+25 e^{4 \alpha }}\right) \delta }{2 \left(1-e^{2 \alpha }\right)}\right)$                    &
\\[4mm] \hline
& & \\[-5mm] 
 $\dfrac{2 \left(1+e^{2 \alpha }+2 \sqrt{1-e^{2 \alpha }+e^{4 \alpha }}\right) \delta }{1-e^{2 \alpha }}$& $\left(0,\dfrac{\left(1+e^{2 \alpha }+2 \sqrt{1-e^{2 \alpha}+e^{4 \alpha }}\right) \delta }{1-e^{2 \alpha }}\right)$ &
\\[4mm] \hline
& & \\[-5mm] 
 $\dfrac{2 \left(1+e^{2 \alpha }-2 \sqrt{1-e^{2 \alpha }+e^{4 \alpha }}\right) \delta }{1-e^{2 \alpha }}$& $\left(0,\dfrac{\left(1+e^{2 \alpha }-2 \sqrt{1-e^{2 \alpha }+e^{4 \alpha }}\right) \delta }{1-e^{2 \alpha }}\right)$ &
\\[4mm] \hline
& & \\[-5mm] 
$\dfrac{\left(5 e^{2 \alpha }-1-\sqrt{25-18 e^{2 \alpha } +9e^{4 \alpha }}\right) \delta }{1-e^{2 \alpha }}$& $\left(-\delta,\dfrac{\left(1+3 e^{2 \alpha }-\sqrt{25-18 e^{2 \alpha }+9 e^{4 \alpha }}\right) \delta }{2 \left(1-e^{2 \alpha }\right)}\right)$                         &
\\[4mm] \hline
& & \\[-5mm] 
$\dfrac{\left(5 e^{2 \alpha }-1+\sqrt{25-18 e^{2 \alpha } +9e^{4 \alpha }}\right) \delta }{1-e^{2 \alpha }}$& $\left(-\delta,\dfrac{(1 +3 e^{2 \alpha } +\sqrt{25-18 e^{2 \alpha }+9 e^{4 \alpha }}) \delta }{2(1- e^{2 \alpha })}\right)$      &
\\[4mm] \hline
& & \\[-5mm]
$\dfrac{2 \left(3+e^{2 \alpha }\right) \delta }{1-e^{2 \alpha }}$ & $\left(2\delta, \dfrac{(1 +3 e^{2 \alpha }) \delta }{1-e^{2 \alpha }}\right)$ & 
\\[4mm]\hline  
& & \\[-5mm]
 $\dfrac{2 \left(1+3 e^{2 \alpha }\right) \delta }{1-e^{2 \alpha }}$ & $\left(-2\delta, \dfrac{\left(3+e^{2 \alpha }\right) \delta }{1-e^{2 \alpha }}\right)$ &
 \\[4mm] \hline 
& & \\[-5mm]
 \multirow{3}{*}{$\dfrac{10 \left(1+e^{2 \alpha }\right) \delta }{1-e^{2 \alpha }}$} & 
 $
 \left(\dfrac{\left(5+5 e^{2 \alpha }+\sqrt{1-22 e^{2 \alpha }+e^{4 \alpha }}\right) \delta }{2(1-e^{2 \alpha })}, \right.
 $
 & \\
 & $
 \quad\quad\left.\dfrac{\left(5+5 e^{2 \alpha }-\sqrt{1-22 e^{2 \alpha }+e^{4 \alpha }}\right) \delta }
 {2 \left(1-e^{2 \alpha}\right)}\right)$
 &
\\[4mm] \hline
 \end{tabular}\\
\caption{The energy spectrum and associated solutions of the Bethe ansatz equations (\ref{specbae}) on the lines (\ref{lines}) for $L=6,\,M=2$. While the pairs $(2\delta,\delta)$, $(\delta,0)$, $(0,-\delta)$, $(-\delta,-2\delta)$ are solutions of (\ref{specbae}) they do not correspond to an eigenvalue of the Hamiltonian, giving rise to a generalised exclusion principle.}
\end{center}
\end{table} 

\section*{Appendix B - Mean field analysis}

In the Appendix we show that a mean-field annalyis in the continuum limit gives the result that the spectrum of the Hamiltonian (\ref{bcs_eqn}) subject to (\ref{gees},\ref{pf}) is real for all values of the coupling parameters. For this reason it is deemed necessary to analyse the model through the exact Bethe ansatz solution.

For a general non-Hermitian Hamiltonian with a real spectrum, where $|\Psi^{\rm r}\rangle$ denotes the right ground-state eigenvector and $\langle\Psi^{\rm l}|$ denotes the left ground-state eigenvector, the ground-state energy can be expressed as 
 $E={\langle \Psi^{\rm l}|H|\Psi^{\rm r}\rangle}/{\langle \Psi^{\rm l}|\Psi^{\rm r}\rangle}$
 provided $\langle \Psi^{\rm l}|\Psi^{\rm r}\rangle\neq 0$.  
When this is the case, for any operator $A$ we {\it define} the ground-state expectation value as 
$\langle A \rangle ={\langle \Psi^{\rm l}|A|\Psi^{\rm r}\rangle}/{\langle \Psi^{\rm l}|\Psi^{\rm r}\rangle}$.
This definition preserves the Hellmann-Feynman theorem.
  
Given a BCS Hamiltonian of the form (\ref{bcs_eqn}), we introduce a (real-valued) chemical potential $\mu$ and set 
\begin{align*}
\Delta^{\rm l}_j &= \sum_{k=1}^L G_{jk}\left<c_{k-}c_{k+}\right>, &
{\Delta}^{\rm r}_k&=\sum_{j=1}^L G_{jk} \left<c_{j+}^\dagger c_{j-}^\dagger\right> 
\end{align*}
to obtain the mean-field approximation
\begin{align*}
{H}_{\rm MF}
&= \sum_{j=1}^L \epsilon_j n_j - \sum_{j=1}^L \Delta^{\rm l}_j c_{j+}^\dagger c_{j-}^\dagger 
- \sum_{k=1}^L {\Delta}^{\rm r}_k c_{k-}c_{k+}   
+  \sum_{j,k=1}^L G_{jk}\left<b_j^\dagger\right>\left< b_k\right>-\mu\left( n-\langle{n}\rangle \right).
\end{align*}
 Setting $
\xi_k=  \epsilon_k -\mu$,
$\mathcal{E}_k = \sqrt{\xi_k^2+\Delta_k^{\rm l}{\Delta}^{\rm r}_k}$, 
by diagonalising ${H}_{\rm MF}$ it is seen that the elementary excitation energies are simply the $\mathcal{E}_k$. 
The entire energy spectrum is real-valued if the ground-state energy is real-valued and the products $\Delta_k^{\rm l}{\Delta}^{\rm r}_k$ are non-negative for all $k$. 

The right mean-field ground state is given by 
$$|\Psi^{\rm r}\rangle=\prod_{k=1}^L(u_k^{\rm r}I +v_k^{\rm r} c^\dagger_{k+}c^\dagger_{k-})|0\rangle$$ 
 where 
${v^{\rm r}_k}/{u^{\rm r}_k}= ({\mathcal{E}_k-\xi_k})/{{\Delta}^{\rm r}_k}={{\Delta}^{\rm l}_k}/({\mathcal{E}_k+\xi_k}) $. 
 Analogously for the left ground state, 
 $$\langle\Psi^{\rm l}|=\langle 0|\prod_{k=1}^L({u}^{\rm l}_kI +{v}^{\rm l}_k c_{k-}c_{k+}),$$ 
 we have
 ${{v}^{\rm l}_k}/{{u}^{\rm l}_k}= ({\mathcal{E}_k-\xi_k})/{{\Delta}^{\rm l}_k}={{\Delta}^{\rm r}_k}/({\mathcal{E}_k+\xi_k})$.   
 Then
$\langle \Psi_{\rm l}|\Psi_{\rm r}\rangle$ is found to be non-zero provided $\Delta^{\rm l}_k{\Delta}^{\rm r}_k\neq 0$ for all $k$.

 Self-consistency requirements impose that 
 \begin{align}
 \Delta_j^{\rm l}
 &= \frac{1}{2}\sum_{k=1}^L G_{jk}  \frac{\Delta^{\rm l}_k}{\mathcal{E}_k}, \label{g1} \\
 {\Delta}^{\rm r}_k 
 &= \frac{1}{2}\sum_{j=1}^L G_{jk}\frac{{\Delta}^{\rm r}_j}{\mathcal{E}_j}, \label{g2} \\
 \langle n \rangle &=L+\sum_{j=1}^L \frac{\mu-\epsilon_j}{{\mathcal E}_j  } .
\label{chempot} 
 \end{align}
 We refer to (\ref{g1},\ref{g2}) as the as the gap equations, and to (\ref{chempot}) as the chemical potential equation. 
 Using these equations allows for the mean-field ground-state energy to be expressed as  

\begin{align*}
{E}_{\rm MF} 
&= \sum_{k=1}^L\epsilon_k\left(1-\frac{\xi_k}{\mathcal{E}_k}\right)-\frac{1}{2} \sum_{k=1}^L \frac{\Delta^{\rm l}_k{\Delta}^{\rm r}_k}{\mathcal{E}_k}
\end{align*}

Note that for a Hermitian Hamiltonian of the form (\ref{bcs_eqn}), $\Delta^{\rm l}_k,\,\Delta^{\rm r}_k$ are a complex conjugate pair, in which case (\ref{g1},\ref{g2}) are equivalent.  
The question remains whether for non-Hermitian Hamiltonians there exist solutions of (\ref{g1},\ref{g2},\ref{chempot}) such that $\Delta_k^{\rm l}{\Delta}^{\rm r}_k$ is real-valued for all $k$. For the choice (\ref{gees}) we next show that this is the case in the continuum limit. 
 We introduce a cut-off energy $\omega$ by setting  $\delta=2\omega/(L-1)$ such that
$\epsilon_1=-\omega$ and $\epsilon_L=\omega$. 
Letting $G_{\pm}=4\omega g_{\pm}/L$, $x=\langle{n}\rangle/(2L)$,  in the continuum limit $L\rightarrow \infty,\,G_{\pm}\rightarrow 0,\,\delta\rightarrow d\epsilon'$ we have (\ref{g1},\ref{g2},\ref{chempot}) assuming the integral equation forms
\begin{align}
\Delta^{\rm l}(\epsilon) &= g_-\int_{-\omega}^\epsilon d\epsilon' \,\frac{\Delta^{\rm l}(\epsilon')}{\mathcal{E}(\epsilon')}
+g_+\int^{\omega}_\epsilon d\epsilon'\,\frac{\Delta^{\rm l}(\epsilon')}{\mathcal{E}(\epsilon')} 
\label{eq2} \\
 {\Delta}^{\rm r}(\epsilon)  &= g_+\int_{-\omega}^\epsilon d\epsilon' \,\frac{{\Delta}^{\rm r}(\epsilon')}{\mathcal{E}(\epsilon')} +
 g_-\int^{\omega}_\epsilon d\epsilon' \,\frac{{\Delta}^{\rm r}(\epsilon')}{\mathcal{E}(\epsilon')}
 \label{eq3}  \\
 x&=\frac{1}{2}+\frac{1}{4\omega}\int_{-\omega}^{\omega} d\epsilon'\,\frac{\mu-\epsilon'}{\mathcal{E}(\epsilon')  } 
\label{eq1} 
 \end{align}   
 where $\mathcal{E}(\epsilon)=\sqrt{(\epsilon-\mu)^2+\Delta^{\rm l}(\epsilon){\Delta}^{\rm r}(\epsilon)}$.
 Differentiating (\ref{eq2}) and (\ref{eq3}) leads to the conclusion 
that $\Delta=\sqrt{\Delta^{\rm l}(\epsilon){\Delta}^{\rm r}(\epsilon)}$ is constant.  
 With this observation the integrals (\ref{eq2},\ref{eq3},\ref{eq1}) can be evaluated to obtain  
\begin{align*}
\mu=\frac{\omega(g_+^{\chi}+g_-^{\chi})(2x-1)}{g_+^{\chi}-g_-^{\chi}},  \quad
\Delta^2=\frac{16\omega^2 g_+^\chi g_-^\chi x(1-x)}{(g_+^\chi-g_-^\chi)^2},
\end{align*}
where $\chi=1/(g_+-g_-)$. Note that $\Delta^2>0$ for all $x,\,g_{\pm}$.
Now  the elementary excitation spectrum is explicit, viz.
$\mathcal{E}(\epsilon)=\sqrt{(\epsilon-\mu)^2+\Delta^2},\, -\omega\leq \epsilon\leq \omega,$
such that $\Delta$ is the gap. We obtain the gound-state energy per fermion as
\begin{align*}
e_{\rm MF}&=\lim_{L\rightarrow\infty} \frac{E_{\rm MF}}{2xL} \\
&=-\frac{1}{8x\omega} \int_{-\omega}^\omega d\epsilon\,\frac{2\epsilon(\epsilon-\mu)+\Delta^2}{\sqrt{(\epsilon-\mu)^2+\Delta^2}} \\
&=-\frac{1}{8x\omega}\left((\omega+\mu)\sqrt{(\omega-\mu)^2+\Delta^2} +(\omega-\mu)\sqrt{(\omega+\mu)^2+\Delta^2}\right). 
\end{align*}
Thus within this mean-field analysis the non-Hermitian Hamiltonian has a real spectrum for all couplings $g_\pm$ and filling fractions $x$. 
In the Hermitian limit 
$g_\pm \rightarrow g$ for which $\chi\rightarrow \infty$, both  $\mu$ and $\Delta$ can be evaluated through use of  
$\displaystyle \exp(x)=\lim_{n\rightarrow \infty} \left(1+{x}/{n}\right)^n$.
In particular for half-filling $x=1/2$ we obtain $\mu=0$ and  
$\Delta={\omega}/{\sinh(1/2g)}$ 
which is in agreement with the classic $s$-wave result obtained in \cite{bcs57} (equation (2.40)).

It might be expected that mean-field results are exact
in the thermodynamic limit (e.g. see \cite{rsd02} for when $g_+$ = $g_-$). It would be very useful
to determine whether or not the energy spectrum is real to leading order, with complex
terms only appearing in lower order corrections. 

\ack Jon Links  and Yao-Zhong Zhang are supported by the Australian Research Council through Discovery Projects DP110101414 and DP110103434 respectively. Amir Moghaddam is supported by an International Postgraduate Research Scholarship and a UQ Research Scholarship.


\section*{References}
\begin{thebibliography}{50}

 
\bibitem{als05} A. Anfossi, A. LeClair, and G. Sierra, {\it The elementary excitations of the exactly solvable Russian doll BCS model of superconductivity}, J. Stat. Mech.: Theor. Exp., P05011 (2005)

\bibitem{b02} R.J. Baxter, {\it Completeness of the Bethe ansatz for the six and eight-vertex models}, J. Stat. Phys. {\bf 108}, 1 (2002).  

\bibitem{bcs57} J. Bardeen, L.N. Cooper, and J.R. Schrieffer, {\it Theory of superconductivity}, Phys. Rev. {\bf 108}, 1175 (1957). 

\bibitem{bfgj12} C. Bender, A. Fring, U. G\"unther, and H. Jones, {\it Preface - Quantum physics with non-Hermitian operators}, J. Phys. A: Math. Theor. {\bf 45}, 440301 (2012).

\bibitem{b31} H. Bethe, {\it Zur Theorie der Metalle}, Z. Physik {\bf 71}, 205 (1931).

\bibitem{cp91} V. Chari and A. Pressley, {\it Yangians and $R$-matrices}, L'Enseignement Math. {\bf 36}, 267 (1990).


\bibitem{dl04} C. Dunning and J. Links, {\it Integrability of the Russian Doll BCS model}, Nucl. Phys. B {\bf 702}, 481 (2004). 

\bibitem{fm01} K. Fabricius B.M. McCoy, {\it Bethe's equation is incomplete for the $XXZ$ model at roots of unity}, 
Stat. Phys. {\bf 103}, 647 (2001).

\bibitem{fs07} P. Fendley and K. Schoutens, {\it Cooper pairs and exclusion statistics from coupled free-fermion chains}, J. Stat. Mech.: Theor. Exp., P02017 (2007).

\bibitem{h91} F.D.M. Haldane, {\it ``Fractional statistics'' in arbitrary dimensions: a generalization of the Pauli principle}, Phys. Rev. Lett. {bf 67}, 937 (1991).

\bibitem{i94} S.B. Isakov, {\it Statistical Mechanics for a class of quantum statistics}, Phys. Rev. Lett. {\bf 73}, 2150 (1994).

\bibitem{ks79} P.P. Kulish and E.K. Sklyanin, {\it Quantum inverse scattering method and the Heisenberg ferromagnet}, Phys. Lett. {\bf 70A}, 461 (1979). 

\bibitem{lrs04} A. LeClair, J.M. Rom\'an, and G. Sierra, {\it Russian doll renormalization group and superconductivity}, Phys. Rev. B {\bf 69}, 020505(R) (2004).

\bibitem{lmz12} J. Links, A. Moghaddam, and Y.-Z. Zhang, {\it Deconfined quantum criticality and generalised exclusion statistics in a non-Hermitian BCS model}, J. Phys. A: Math. Theor. {\bf 45}, 462002 (2012).

\bibitem{p96} A.P. Polychronakos, {\it Probabilities and path-integral realization of exclusion statistics}, Phys. Lett. B {\bf 365}, 202 (1996).

\bibitem{rsd02} J.M. Rom\'an, G.Sierra, and J. Dukelsky, {\it Large-$N$ limit of the exactly solvable BCS model: analytics versus numerics}, Nucl. Phys. B {\bf 634}, 483 (2002). 

\bibitem{sbsvf04} T. Senthil, L. Balents, S. Sachdev, A. Vishwanath, and M.P.A. Fisher, {\it Quantum criticality beyond the Landau-Ginzburg-Wilson paradigm}, Phys. Rev. B {\bf 70}, 144407 (2004).

\bibitem{svbsf04} T. Senthil, A. Vishwanath, L. Balents, S. Sachdev, and M.P.A. Fisher, {\it Deconfined quantum critical points}, Science {\bf 303}, 1490 (2004).

\bibitem{tf79} L. A. Takhtadzhan and L.D. Fadeev, {\it The quantum method of the inverse problem and the Heisenberg $XYZ$ model}, Russ. Math. Surv. {\bf 34}, 11 (1979).

\bibitem{r63} R.W. Richardson, {\it A restricted class of exact eigenstates of the pairing-force Hamiltonian}, Phys. Lett. {\bf 3}, 277-279 (1963).

\end{thebibliography}
\end{document}